\documentstyle[12pt,epsfig]{article} 
\begin{document}
\begin{center}
{\bf The Baryon Wilson Loop Area Law in QCD}\\[.2in]
John M. Cornwall*\\[.2in]
{\it Physics Department, University of California\\
405 S. Hilgard Ave., Los Angeles Ca 90095-1547}\\[.2in]
{\bf Abstract}
\end{center}

There is still confusion about the correct form of the area law for the
baryonic Wilson loop (BWL) of QCD.  Strong-coupling (i.e., finite lattice spacing in lattice gauge theory) approximations suggest
the form $\exp[-KA_Y]$, where $K$ is the $q\bar{q}$ string tension
and $A_Y$ is the global minimum area, generically a three-bladed area
with the blades joined along a Steiner line ($Y$ configuration).  However,
the correct answer is $\exp[-(K/2)(A_{12} + A_{13} + A_{23})]$, where, e.g.,
$A_{12}$ is the minimal area between quark lines $1$ and $2$ ($\Delta$
configuration). This second answer was given long ago, based on certain
approximations, and is also strongly favored in lattice computations.  In the present work, we derive the $\Delta$ law from the
usual vortex-monopole picture of confinement, and show that in any case
because of the $1/2$ in the $\Delta$ law, this law leads to a larger value for the BWL 
(smaller exponent) than does the $Y$ law.  We show that the three-bladed strong-coupling surfaces, which are infinitesimally thick in the limit of
zero lattice spacing, survive as surfaces to be used in the non-Abelian
Stokes' theorem for the BWL, which we derive, and lead via this Stokes'
theorem to the correct $\Delta$ law.  Finally, we extend these considerations,
including perturbative contributions,
to gauge groups $SU(N)$, with $N > 3$.\\[1.3in]
UCLA/96/TEP/15 \\
May 1996\\
\footnoterule
\noindent * Electronic address:  Cornwall@physics.ucla.edu
\newpage

\section{Introduction}

QCD is a theory more than twenty years old, yet certain questions of
fundamental principle seem still to have no definitive answer.  Among these
is the question of the correct form of the area law for the baryonic 
Wilson loop, or BWL for short (and its analog for $SU(N)$ gauge groups with $N > 3$).

The BWL is defined as:

\begin{equation} BWL \equiv \frac{1}{6} \epsilon_{abc} \epsilon_{a'b'c'}
U(x, y; 1)_{aa'} U(x, y; 2)_{bb'}U(x, y; 3)_{cc'}
\end{equation}
where the quark lines, labeled $1, 2, 3$, emerge from the point $x$ and
rejoin the vacuum at point $y$.  The $U's$ are defined by, for example:
\begin{equation} U(x, y; 1)_{aa'} \equiv (P \exp \int_{\Gamma(1)} dz \cdot A(z))_{aa'}
\end{equation}
where $P$ stand for path ordering, the integral runs from $x$ to $y$, and
$\Gamma(1)$ is the path from $x$ to $y$ for quark line $1$ (see Fig. 1).  We define the vector potential $A(z)$ as:
\begin{equation} A_{\mu}(z) = \frac{g\lambda_j}{2i}A_{\mu}^j(z)
\end{equation}
in terms of the usual component potential $A_{\mu}^j(z)$ and Gell-Mann
matrices $\lambda_j$; $g$ is the coupling constant.
 
    There are only two contenders for the form of the BWL area law; both
were given early on and continue to be discussed.  The first is\cite{xa, dm, cjp, ip, kn} 
\begin{equation} \langle BWL  \rangle = \exp[-KA_Y]       %4%
\end{equation} 
and the second is\cite{c77}
\begin{equation} \langle BWL \rangle = \exp[-KA_{\Delta}/2].
\end{equation}
In both (4) and (5), $K$ is the $q\bar{q}$ string tension; in (4), $A_Y$
is the three-bladed area running from each quark line to a central Steiner
line, which generically exists to define the global minimum area (see Fig. 2),
and in (5),
\begin{equation} A_{\Delta} \equiv A_{12} + A_{13} + A_{23}
\end{equation}
 where $A_{ij}$ is the minimal area spanning quark lines $i$ and $j$
(see Fig. 3).  We call the area law (4) the $Y$ law, and (5) we call the
$\Delta$ law.  It is easy to understand the normalization of the exponent:
When lines $1$ and $2$, say, are made to coincide, then the corresponding
quarks act like a single antiquark, which must show the usual $q\bar{q}$
area law, based on the minimal area $A$ spanning line $3$ and the effective
antiquark.  But when lines $1$ and $2$ coincide, the coefficients of $K$ in
both area laws reduce to $A$.

The author's previous argument\cite{c79} for the $\Delta$ law was based
on a vortex condensate model discussed below, but it used some
approximations which are in fact completely unnecessary and which will
not be used here.  Furthermore, some vital technical details were omitted,
notably concerning a non-Abelian Stokes' theorem for the BWL, which we supply
and use in the present work.

Given that the early work on the BWL was based on approximations and 
intuitive insight, can we today say which of the laws (4) or (5) is correct?  We will show here that the $\Delta$ law in
(5) is, based on the continuum version\cite{c79} of confinement
via a vortex-monopole condensate.  These vortices were invoked by
't Hooft\cite{th79}, and developed on the lattice by Mack and Petkova\cite{mp}.
These lattice considerations are not of the strong-coupling (finite lattice
spacing) type, to which we will come in a minute; they are intended to apply
to the weak-coupling or continuum limit of zero lattice spacing.  
Tomboulis\cite{t93} has more recently given some rigorous results on these
lattice developments (unfortunately only for $SU(2)$, where there is no
BWL), confirming that confinement can only come from a condensate of
vortices\footnote{Tomboulis also gives a number of references to other works
developing the vortex picture on the lattice.}.  In our arguments for the
$\Delta$ law (5), the coefficient $K/2$ is derived by simultaneously deriving the $q\bar{q}$ area law and the BWL area law from the vortex condensate.

We state our arguments about the BWL and the vortex condensate in terms of
a specific realization of the condensate, based on dynamical and gauge-invariant generation of a gluon mass, dynamically consistent because the gluon mass
vanishes at short distances\cite{c79,c82,cy,la}.  While one might doubt the
accuracy of specific quantitative predictions of any condensate model, our
results concerning the BWL depend only on general features of the vortex
condensate picture.  It is merely for concreteness of exposition that we 
choose to use the picture of vortices driven by a dynamical gluon mass.
This independence of details of the vortex condensate holds as long as the
BWL is large, in the sense that all scale lengths of the quark lines in the
BWL (length, distance of closest approach to itself or other lines, curvature radius, torsion length, etc.) are large compared to the
scale length $\Lambda^{-1}$ of QCD, and as long as we are only interested in 
the area law and not perimeter law corrections.  The general features are:
\begin{enumerate}
\item QCD field strengths are short-ranged.
\item QCD gauge potentials have a pure-gauge long-range part (if there
is no long-range part, there is no area law, only a perimeter law).
\item  The magnetic fluxes of the vortices (which realize points 1, 2 above)
lie in the center of the gauge group.
\item  Because of the finite correlation length $\Lambda^{-1}$ of QCD, 
distinct vortices are statistically independent.  Indeed, parts of a single
vortex which are separated by a distance large compared to $\Lambda^{-1}$
are uncorrelated.
\end{enumerate}
If the BWL is large, then only the long-range pure gauge part of the
potential need be taken into account, which greatly simplifies the argument.
Confinement and the consequent area laws become a matter of counting Gauss
linking numbers\cite{c79,cy}, since both the $q\bar{q}$ Wilson loop and the
BWL can be expressed in terms of standard Gauss linking integrals.  These
integrals can be interpreted by using the non-Abelian Stokes' theorem
to convert them to integrals counting the (signed) intersections of the
vortices with surfaces spanning the $q\bar{q}$ Wilson loop or the BWL.
We will discuss this Stokes' theorem for the BWL below.  Following this
line of thought leads immediately to the form (5) for the BWL area law.    

The alternative $Y$ law in (4) is usually argued for\cite{xa, dm, cjp, ip, kn}
on the basis of lattice strong-coupling arguments, with the lattice spacing
$a$ kept finite and $O(\Lambda^{-1})$, where $\Lambda$ is the QCD mass scale,
or intuitive remarks are made to the effect that $A_Y$ is the minimum area.
However, we will show very easily that while $A_Y \leq A_{\Delta}$, in fact
\begin{equation}  A_Y \geq \frac{1}{2}A_{\Delta}.  
\end{equation}
This shows that the exponent in (5) is less than that in (4), so any
intuitive argument about minimum areas should favor the $\Delta$ law\footnote{Equality is reached in two dimensions, that is, when all three
quark lines lie in a plane.}.

We note that lattice
calculations of the BWL have always favored the $\Delta$ law\cite{sw,tes}
and the last-cited authors claim that there is no evidence for the $Y$ law.

Given that the $\Delta$ law is correct, what has happened to the strong-coupling surfaces that underlie the $Y$ law?  These survive, in a sense, as mathematical surfaces of infinitesimal thickness (lattice-space thickness) which are of
(color) electric character.  By this we mean that these surfaces are bounded
by quark Wilson lines.  The infinitesimally-thick surfaces are dual to closed
magnetic surfaces which comprise the condensate of vortices in the vacuum---
magnetic because a static vortex has only short-range color magnetic fields.
Confinement for a conventional ($q\bar{q}$) Wilson loop is described\cite{c79,cy,mp,t93} as linkage of the Wilson loop with the closed
surface\footnote{In dimension $d$ a Wilson loop can link with a surface of
codimension 2, that is, a point in two dimensions, another closed loop in
three dimensions.}.   To describe this linkage one needs to use a non-Abelian
version of Stokes' theorem appropriate for the BWL, by means of which one
converts the line integral in the Wilson loop to a surface integral, and
notes that the surface integral is an intersection number of the magnetic
surface for the vortex and an electric surface spanning the Wilson loop.
Because of the intimate connection with Stokes' theorem, we henceforth call
these electric surfaces by the name of Stokes surfaces.  They are precisely
the surfaces invoked in the strong-coupling approximation.
The necessary non-Abelian Stokes' theorem for the $q\bar{q}$ loop has been
long known\cite{br,ar,fe,fgk}, but the author is not familiar with an
analogous discussion for the BWL.  We give here the non-Abelian Stokes' 
theorem for the BWL, which is a quite elementary variation on the usual
non-Abelian Stokes' theorem.  The interesting feature of the BWL theorem is
that the Stokes surface is a three-bladed ($N$-bladed) surface for $SU(3)$
($SU(N)$) bounded by
the BWL, with the blades running from the quark lines to a central line; that is, for $SU(3)$ it is just the strong-coupling surface of Fig. 2.  Just as for the $q\bar{q}$ Stokes' theorem, any surface of the
proper topological type may be used and the BWL is quite independent of the
choice of surface, including the choice of the central line where the blades
meet.

At first sight, this last feature is surprising.  A simple Abelian vortex
linked to quark line 1 and penetrating only one leaf of the three-bladed surface (see Fig. 4) has a link number of 1, while if the Steiner line is pulled inside the vortex it has a link number of -2 (see Fig. 5).  But it turns out that
the only thing that matters is the link number mod 3, so these two situations
are equivalent.  Of course, this mod 3 dependence is to be expected, in view
of the fact that confinement involves the center of the group, in this case
$Z_3$, and that magnetic fluxes are quantized in units of $2\pi /3$.

The next question one might ask is how these results are generalized to
$SU(N),\;N>3$.  The answer is, as we will indicate, that the analog of the
$\Delta$ law holds, with the result\footnote{Of course, for $N$ even the
BWL does not describe a baryon.}               %8%
\begin{equation}  \langle BWL \rangle = \exp[-KA_{\Delta}/(N-1)]\end{equation}
where
\begin{equation} A_{\Delta} \equiv \sum_{i<j}A_{ij}  \end{equation}
and $A_{ij}$ is the minimal area between legs $i$ and $j$.  One can also show
the analogous inequality to (7)
\begin{equation} A_Y \geq \frac{A_{\Delta}}{N-1}  \end{equation}
where now $A_Y$ means the generic minimum Steiner surface, which has
$N-2$ Steiner lines where three surfaces meet.  Note that for all $N$ the factor of $1/(N-1)$ in the 
$\Delta$ law (8) also occurs for lowest-order gluon exchange (plus radiative corrections to the one-gluon potential and certain other two-body graphs),
so that the lowest-order two-body potential is $1/(N-1)\Sigma_{i<j}V_{ij}$ where $V_{ij}$ is the $q\bar{q}$ one-gluon potential:
\begin{equation} V_{ij} = -\frac{g^2C_F}{4\pi r_{ij}}.  \end{equation}        
 (Here
$C_F$ is the quark Casimir, of $O(N)$.)Note that these two-body forces have, so to speak, the topological character of the $\Delta$ law, as well as its numerical coefficient.  Similarly, a cross-section of the generic Steiner
surface reminds one of perturbative graphs with $N-2$ three-gluon vertices.

In connection with perturbative contributions, where there is interest not only for QCD but also for large $N$\cite{th74,wi}\footnote{Large-$N$ studies of baryons have been
modernized recently; see Refs. \cite{djm,cgo,lmw} which cite other
recent references.},  recall that all graphs for the BWL are individually non-leading at
large $N$.  However, the sum of all possible graphs of a given type may
be leading; for example, there are $N(N-1)/2$ one-gluon exchange
graphs, each of strength $g^2C_F/(N-1)$, so that the sum is $O(N)$. But
many perturbative contributions which are formally leading in this sense vanish identically; in particular,
it was shown long ago\cite{c77,ff} that in $SU(3)$ the lowest-order graph with one gluon line on each leg meeting at one three-gluon vertex
vanishes identically for group-theory reasons. We will discuss some of these graphs in Section 4 for $N>3$, where we show that for any $N$ the generalization of the above graph, with three gluons attached to different quark lines and meeting
at a three-gluon vertex, vanishes identically via a simple symmetry argument.  Some other higher-order perturbative graphs with three-gluon vertices vanish for the same reasons, and it may be of interest to note that those gluon-tree graphs with the
topology of a cross-section of the generic minimal Steiner surface give zero. 
\section{Area Laws}
\subsection{Review of the $q\bar{q}$ Wilson Loop}

Before considering the BWL area law, we set the stage by briefly reviewing
the vortex-condensate picture of confinement in the usual $q\bar{q}$ Wilson
loop.  The condensate is formed\cite{c79,cy} from vortices centered on simple
closed surfaces\footnote{There are vortices described by non-simple closed surfaces where three blades meet along a line; in $SU(3)$, the blades are
associated with the Lie-algebra structures diag(1/3,1/3,-2/3), diag(1/3,-2/3,1/3), and diag(-2/3,1/3,1/3), one for each blade, generalizing (12).  Note that the sum of these generators is zero.  In $d=3$ the
simplest such vortex looks just like the BWL of Fig. 1 itself.  Although these truly non-Abelian vortices may lead to interesting knot-classification problems,   they add nothing new to
our discussion of confinement\cite{c79}.}  of codimension 2; in four Euclidean dimensions, a specific
realization of the vortex potential is:
\begin{equation} igA_{\mu}(x) = 2\pi Q \epsilon_{\mu \nu \alpha \beta}
\partial_{\nu} \int d\sigma_{\alpha \beta}[\Delta_M(x-z) - \Delta_0(x-z)]
\end{equation}
where
\begin{equation} Q = \rm{diag} (1/3, 1/3, -2/3) \end{equation}
and $d\sigma_{\alpha \beta}$ is the element of surface on the surface
$z(\sigma, \tau)$:
\begin{equation}  d\sigma_{\alpha \beta} = \dot{z}_{[\alpha}z'_{\beta ]}
d\sigma d\tau
\end{equation}
in standard notation.  In (12), $\Delta_M$ is a free propagator of mass $M$
and $\Delta_0$ is a massless free propagator; the mass $M$ is a dynamically-
generated mass\cite{c79,c82,la,cy} of $O(\Lambda)$.  This mass is generated gauge-invariantly
and without symmetry breaking, and its kinematical description requires
the massless term in (12), which is actually a pure gauge term as one can
check directly using Stokes' theorem.  The normalization is chosen so that
parallel transport around a closed path (which may link the closed vortex
surface but is far from it) gives an element of the gauge group lying in the
center $Z_3$; of course, only the massless term contributes in this transport
if the closed path and vortex surface are separated by distances $\gg M^{-1}$.
If the closed path and the vortex surface are not linked only the identity
element of $Z_3$ can occur.
Any regular gauge transform of (12) is also allowed, in particular, the
diagonal elements of $Q$ can be permuted.

Consider a Wilson loop which is large in the sense described earlier; for
such a loop, the contribution of a vortex to the area law for a vortex
which never gets close to the loop can be found from only the $\Delta_0$
term in (12).  The massive term only contributes to perimeter-law corrections
from vortices within a distance $M^{-1}$ of the Wilson loop.  
The expectation value of such a Wilson loop is:
\begin{eqnarray} \langle W \rangle & = & \frac{1}{3}\langle Tr P \exp \oint dx \cdot A(x)
\rangle \\
& = & \langle \exp [2\pi i L/3] \rangle \nonumber \end{eqnarray}
where 
\begin{equation} L = \sum L_i \end{equation}
is the sum of the Gauss linking numbers of vortex $i$, as given by
\begin{equation} L_i = \oint dx_{\mu} \oint d\sigma_{\alpha \beta}(i)
\epsilon_{\alpha \beta \mu \nu}\partial_{\nu}\Delta_0(x-z(i)).
\end{equation}
Converting the $x$ integral to a surface integral by Stokes' theorem shows
that this linking number is an intersection number of the vortex surface
with a surface bounded by the Wilson loop; which particular surface is chosen
is immaterial.  The total number $N$ of vortices linked to the loop is clearly
proportional to some area $A$ associated with the loop, and the coefficient of
proportionality is the density of vortices $\rho$ per unit area:  $N=\rho A$.  Since variation of
the area $A$ can lead to no change in $\langle W \rangle$, this area must be
the minimum spanning area.

Now we invoke the above-stated assumption that the linking numbers $L_i$ of distinct
vortices are independent of one another.  Because the area $A$ is large, $N$ is
also large and we invoke the central limit theorem to conclude that $L$, the
sum of a large number of independent random variables $L_i$, has a Gaussian
distribution with average value $\langle L \rangle = 0$ and that
\begin{equation} \langle \exp[2\pi iL/3] \rangle =  \exp[
-\frac{1}{2}(2\pi /3)^2 \rho A \langle L_i^2 \rangle] \equiv e^{-KA} \end{equation}
where $K$ is the mesonic string tension.

For purposes of studying the BWL area law we do not need to know the density
of vortices or how exactly this density is related to $K$; we simply need
to know that 
\begin{equation} \langle W \rangle = \langle \exp[2\pi iL/3] \rangle
= \exp[-const.\langle L^2 \rangle] \end{equation}
and that
\begin{equation} \langle L^2 \rangle = \langle \sum L_i^2 \rangle\sim A.\end{equation}
We now go on to the BWL.

\subsection{The BWL Area Law}
It is not apparent at first glance that the BWL area law can be expressed
in terms of standard Gauss linkages, because there are no obvious directed
closed loops of the standard Wilson-loop type.  If the problem were 
Abelian there would indeed be no simple solution, but of course it is not.
Nonetheless we can define a concept of linkage and in fact reduce the
problem to one of standard Gauss link numbers.

We begin by observing that, because we save only the pure-gauge part of
the vortices, we can define the notion of linking a vortex with the BWL.
Consider the open line integral $U(x, y; 1)$ from $x$ to $y$ (see Fig. 1) for 
quark line 1, as defined in equation (2).  When the vector
potential is pure gauge $U$ can be expressed in the form
\begin{equation} U(x,y;1) = V(x)V^{-1}(y)  \end{equation}
where $V$ is a local $SU(3)$ matrix.  We now say that a vortex is unlinked
to the BWL if $V(x)V^{-1}(y)$ is the same for all three quark lines.  If $V$
were a regular gauge transformation this would be automatic, since $U$ in
(21) would then be path-independent.  For the singular gauge parts we
encounter in vortices this is not so, because parallel transport around
a closed path can lead to a non-trivial element of $Z_3$, which might
be different for different lines; it is clear from the above discussion of
the usual Wilson loop that this is how linkage numbers are generated.  But
when all three lines give the same value for $U$, the elementary identity
\begin{equation} \frac{1}{6} \epsilon_{abc} \epsilon_{a'b'c'}
U_{aa'}U_{bb'}U_{cc'} = \det U = 1 \end{equation}
shows that there is no contribution from such a vortex to the BWL area law.

We can just as easily define the notion of a vortex with a simple link to,
e. g., line 1, as in Fig. 4 (ignore the Stokes surfaces of this figure for
now), in which case $U(1)$ will not be the same as $U(2)$ and $U(3)$, but
these latter two $U$'s have the same value $S$.  As one's intuition suggests, this is the same configuration as a vortex linked to lines 2 and 3, but with the opposite link number (two quarks equal an antiquark, in the BWL).  To show this, use a variant of (22) to find\cite{c79} 
\begin{equation}  \frac{1}{6} \epsilon_{abc} \epsilon_{a'b'c'}
S_{aa'}S_{bb'} = S^{-1}_{c'c}   \end{equation}
so we do not need a separate definition of a simple vortex link to two lines.

At this point it is convenient, but not necessary, to return to the model
of vortices described in connection with the usual Wilson loop, and described in
equations (12-14).  We easily find
\begin{eqnarray}  BWL & = & \frac{1}{6}\epsilon_{abc} \epsilon_{a'b'c'}
(\exp2\pi iQ\Lambda(1))_{aa'}(\exp2\pi iQ\Lambda(2))_{bb'}
(\exp2\pi iQ\Lambda(3))_{cc'}\nonumber\\
& = & \frac{1}{3}\exp[ \frac{2\pi i}{3}(\Lambda (1) + \Lambda (2) -
2\Lambda (3)] + c.p.   
\end{eqnarray}
where the $\Lambda$'s are defined as Abelian path integrals, possibly
path-dependent:
\begin{equation} \Lambda(i) = \int_{\Gamma(i)} dx\cdot A(x) 
\end{equation}
and $c.p.$ stands for cyclic permutations.  

It is clear from (24) that each term of this equation can be written in
terms of conventional loop integrals:
\begin{equation} \Lambda(i) - \Lambda(j) \equiv I_{ij} \equiv
\oint_{\Gamma(i) - \Gamma(j)} dx \cdot A(x)  \end{equation}
with the contour $\Gamma(i) - \Gamma(j)$ oriented to run in the direction
shown in the figures on line $i$, but in the opposite direction to the 
figures on line $j$.  In other words, one has a conventional $q\bar{q}$
contour.  Of course, each integral $I_{ij}$ is a Gauss link integral, so
we have achieved the purpose of expressing the BWL in terms of such
integrals:
\begin{equation}  BWL  = \frac{1}{3}\exp(2\pi i/3)
(I_{13} + I_{23}) + c.p.
\end{equation}

We now define link numbers for elementary linkages in a slightly different way from usual.
The link numbers so defined are called $L(i)$, where the $i$ refers to
the quark line, not any particular vortex\footnote{This notation is to be
distinguished from $L_i$ introduced in (16), which refers to the link number of
a single vortex; for the BWL, $L(i)$ refers to the sum of all vortex link numbers for the quark line $i$.  $L(i)$ is a sum of a large number of link
numbers of statistically-independent individual vortices}.  The first step in the
definition is to write 
\begin{equation} I_{ij} \equiv L(i) - L(j) \end{equation}
A vortex has an elementary linkage to line $i$ if 1) the other two lines can
be continuously deformed (still with ends at $x$ and $y$ in the figures) to coincide with each other without
ever crossing the vortex, and 2) the vortex is then linked to the effective
$q\bar{q}$ vortex formed by line $i$ and the other two lines, with the orientation determined by that of line $i$. Then the elementary link number 
$L(i)$, with sign, is
defined as usual.  For example, the vortex in Fig. 4 has link number +1
with line 1.  More complicated linkages can, in general, be reduced to
simple linkages by reconnecting the vortex itself without changing the
so-defined link numbers, that is, without changing $I_{ij}$, and in such a
way that the elementary link numbers are statistically independent, even if
coming from a single vortex with various twists and writhes along its length.

With this definition, we have for the BWL:
\begin{equation}  BWL = \frac{1}{3}\exp(2\pi i/3)(L(1) + L(2) - 2L(3)) + c.p.
\end{equation}  However, because $L(i)$ is an integer, we may write this
as
\begin{equation} BWL = \frac{1}{3}\exp(2\pi i/3)(L(1) + L(2) + L(3))
\end{equation} 
which shows, as expected, the symmetry under exchange of the quark lines.

Now we invoke the assumed independence of the linking numbers from
distinct vortices (after the above-mentioned process of vortex reconnection,
if necessary), in forming the expectation value $\langle BWL \rangle$
as an average over vortex configurations:
\begin{equation}  \langle L(i)L(j) \rangle = \langle L(i)^2 \rangle
\delta_{ij} \end{equation}
\begin{equation} \langle BWL \rangle = \exp[-\frac{1}{2}(2\pi/3)^2
\sum \langle L(i)^2 \rangle ]    \end{equation}

It only remains to compare this result with that for the $q\bar{q}$ loop
formed on lines $i$ (going up, as in the figures) and $j$ (going down)
as given in terms of link numbers in (18), where the link number called $L$
there is precisely the $I_{ij}$ of (26):
\begin{eqnarray} \langle W_{ij} \rangle & = & \exp[-\frac{1}{2}(2\pi/3)^2
\langle (L(i) - L(j)^2 \rangle \nonumber\\
& = & \exp[-\frac{1}{2}(2\pi/3)^2\langle L(i)^2 + L(j)^2 \rangle     %33%
\nonumber\\
& =& \exp[-KA_{ij}]
\end{eqnarray}
where in the second equality we used the assumption of independence as in (31) and in the last equality we have made use of (20) which states that the
total link number is proportional to the minimal area $A_{ij}$ spanning 
quark lines $i$ and $j$.  By adding the three equations ($(i,j) =(1,2),(1,3),(2,3)$)
\begin{equation} \langle L(i)^2 \rangle + \langle L(j)^2 \rangle
= const. A_{ij}   \end{equation}
we find
\begin{equation}  \sum \langle L(i)^2 \rangle = const. \frac{1}{2}(A_{12} +
A_{13} + A_{23})  \end{equation}
and finally by comparing to the usual Wilson-loop area law in (19), (20) we find the
final result:
\begin{equation} \langle BWL \rangle = \exp[-(K/2)(A_{12} +
A_{13} + A_{23})]  \end{equation}                              %36%
which is the $\Delta$ law claimed in equations (5), (6).

To close this Section we give the simple proof, just a triangle inequality
for areas, that $A_Y \geq A_{\Delta}/2$.  In Fig. 4, define the areas
$A_{0i},\;i=1,2,3$ as the areas spanning quark line $i$ and the central
line 0.  The areas $A_{ij}$ are minimal for the loops formed from quark
lines $i$ and $j$, so
\begin{equation} A_{0i} + A_{0j} \geq A_{ij}\;\;(i\neq j).  \end{equation}
Add these three equations, divide by two, and use the definition (6) of $A_{\Delta}$ plus $A_Y = \Sigma A_{0i}$ to find the needed inequality. 

\section{Non-Abelian Stokes' Theorem for the BWL}

Let us interpret some of the steps of the above discussion in light of
the non-Abelian Stokes' theorem (NAST) for the BWL, proven below.
This theorem states that in the BWL expression (1) each $U$ occurring there
can be replaced by an integral of the type (e. g., for line 1)
\begin{equation} U(x,y;1) \rightarrow P\exp \int_{\Sigma(01)}    %38%
d\sigma_{\mu \nu}(z)\hat{U}(xz)G_{\mu \nu}\hat{U}^{-1}(xz)
\end{equation}
where $\Sigma(01)$ is any surface spanning quark line 1 and a central line
running from $y$ to $x$ which we call line 0 (the dotted line in Fig. 4).
Any set of three surfaces and corresponding central line may be used.  The
$\hat{U}$'s are parallel-transport integrals of a type discussed below, and
$G_{\mu \nu}$ is the usual field strength.  For the special vortex whose
explicit form is given in (12-14), these $\hat{U}$'s commute with 
$G_{\mu \nu}$ but that is not the point here; the point is the nature of
the three-bladed Stokes surface shown in Fig. 4.  The interpretation of the
loop integrals in (26) as linking numbers depends on using Stokes' theorem
for such a surface.  The explicitly-written term in (27) contains---in this sense--- 
contributions from surfaces spanning quarks lines 1 and 3, and lines 2 and 3,
but not lines 1 and 2.  This is consistent with the NAST for the BWL, by choosing the central line of the Stokes surface
to coincide with line 3.  Similarly, the other two contributions in (27)
have the central line chosen to coincide with line 1 or line 2.

First recall the NAST for the usual Wilson
loop\cite{ha,br,ar,fe,fgk}\footnote{Halpern\cite{ha} shows how to fix the gauge so that the line integrals $\hat{U}$ in equation (38) can be replaced by unity, thereby reducing the NAST to its Abelian counterpart.}.  It says that the Wilson loop
\begin{equation} W \equiv \frac{1}{3}Tr P\exp \oint dx \cdot A(x) \end{equation}
can be written
\begin{equation} W = \frac{1}{3} Tr P\exp \int_{\Sigma}d\sigma_{\mu \nu}(z)\hat{U}(oz)G_{\mu \nu}\hat{U}^{-1}(oz).   \end{equation}
Here
\begin{equation} \hat{U}(oz) = \exp \int_o^z dx \cdot A(x)   \end{equation}%41%
is integrated along lines originating at any point $o$ on the loop and
ending at the surface point $z$.  The appropriate paths and the ordering
prescription can be written explicitly, but for our purposes it is enough
to consult the kind of figure drawn by Fishbane, Gasiorowicz, and Kaus
\cite{fgk} and shown here as Fig. 6.  The original square Wilson loop is formed 
from eight segments (labeled 1-8), each segment corresponding to a parallel-
transport integral along that segment.  The original square is subdivided
into four, the first step in dividing the loop into infinitesimal plaquettes,
and the original path is replaced by the one shown in Fig. 6.  It is easy
to see that all the added line integrals cancel each other, so there is no
change in the value of the Wilson loop.  By continuing this process of
plaquette development, the original line integral is turned into a surface
integral of the type (40), with the $\hat{U}$'s defined by reference to
Fig. 6 and its further subdivision into plaquettes.

Now consider the corresponding process for the BWL, a square version of which
is shown in Fig. 7.  We wish to subdivide each three-legged segment into
plaquettes, much as in Fig. 6.  To do so, introduce a central line
(labeled 0 in Fig. 4, or 7-8 in Fig. 8) and subdivide the three squares so
formed just as in Fig. 6, and shown in Fig. 8.  All the extra line integrals
cancel as before, except for the three lines marked 7-8, running from
$y$ to $x$.  As a result, the BWL takes the form
\begin{equation} BWL = \frac{1}{6} \epsilon_{abc} \epsilon_{a'b'c'}
U(x, y; 1)_{ad} U(x, y; 2)_{be}U(x, y; 3)_{cf}U(78)_{da'}U(78)_{eb'}
U(78)_{fc'}                                                           %42%
\end{equation}
where $U(78)$ is the line integral along the central line from $y$ to $x$,
but the original $U(x,y;i)$ can be expressed as surface integrals in the limit
of infinite subdivision into plaquettes.
Once again we use a determinantal argument of the type given in (22), (23):
\begin{equation}  \epsilon_{a'b'c'}
 U(78)_{da'}U(78)_{eb'}
U(78)_{fc'}  = \epsilon_{def}    \end{equation}
which, substituted in (42), shows that the $U(78)$'s do not contribute.  

It is clear that the choice of central line and spanning surfaces is
immaterial to the value of the BWL, just as for the usual Wilson loop.
Written in continuum form, the BWL NAST has already been given in equation (38), where the paths involved in constructing
the $\hat{U}$ path integrals can be read off from Fig. 8.

\section{N $>$ 3 and Perturbative Contributions}

Perhaps the main reason for studying $N>3$ is to make contact with
large-$N$ arguments\cite{th74,wi,djm,cgo,lmw}.

For $N > 3$ the BWL is of the form (1) with $N$ $U$'s going from $x$ to
$y$ as in Fig. 1, and $\epsilon$ symbols of appropriate dimensionality.
The NAST for the BWL has one central line, as in Fig. 4 or Fig. 8, and
the obvious analog of (43) for $SU(3)$ holds; the Stokes surface is then
$N$-bladed, with the blades meeting along the central line.  Note that this
is not the same as the generic minimal surface, which has $N-2$ Steiner lines
where three surfaces meet.  

There are two candidates\footnote{One can guess at many more candidates; for
example, Witten\cite{wi} has speculated on $N$ surfaces meeting along a single
line, which is in fact the Stokes surface for general $N$.  But this sort of
configuration neither comes from the vortex condensate picture nor from an
argument about minimal areas.}, $Y$ and $\Delta$, for the area law, as given in
equations (8-10), which we repeat for convenience:
\begin{equation} \langle BWL  \rangle = \exp[-KA_Y]       %44%
\end{equation} 
\begin{equation}  \langle BWL \rangle = \exp[-KA_{\Delta}/(N-1)]\end{equation}
\begin{equation} A_{\Delta} \equiv \sum_{i<j}A_{ij}  \end{equation}
In (43), $A_Y$ stands for the global minimum area, generically that of a
surface with $N-2$ Steiner lines where three surfaces meet.  As for $N=3$, the normalization is set by the requirement that when $N-1$ quark lines coincide they act like an antiquark and the resulting area law is the $q\bar{q}$ string tension.

We derive the $\Delta$ law with the vortex-condensate picture, as before.  The vortex-
condensate argument is essentially unchanged (with substitution of $N$ for 3
in various places) up to equation (34), which relates the link numbers
$L(i)$ to the minimal areas $A_{ij}$ spanning quark lines $i$ and $j$; these
lines need not be adjacent:
\begin{equation} \langle L(i)^2 \rangle + \langle L(j)^2 \rangle
= const. A_{ij}   \end{equation}
There are $N(N-1)/2$ of these equations, and the sum of them all contains
each $\langle L(i)^2 \rangle$ $N-1$ times.  Then this sum divided by
$N-1$ yields:
\begin{equation} \Sigma \langle L(i)^2 \rangle = const.\frac{1}{N-1}
\sum_{i<j} A_{ij}   \end{equation}
The factor $N-1$ in this equation enters the $\Delta$ law (45) just as in
Section 2.  

We show that                                                       
\begin{equation} A_Y \geq \frac{A_{\Delta}}{N-1}  \end{equation}
explicitly for $N=4$; the idea will then become clear for all $N$.
Fig. 9 shows a cross-section of an $N=4$ BWL, with  the cross-section of the quark lines shown as numbered points connected by a Steiner surface, whose
cross-section is the lines shown, and two Steiner lines A, B.
Then the global minimum area is
\begin{equation}  A_Y = A_{1A} + A_{2A} + A_{3B} + A_{4B} + A_{AB}.
\end{equation}   
There are 6 inequalities expressing the fact that the areas $A_{ij}$
are minimal areas for quark lines $i,j$:
\begin{eqnarray}  A_{1A} + A_{2A} & \geq & A_{12}\\
A_{2A} + A_{3B} + A_{AB} & \geq & A_{23},\;etc. \nonumber \end{eqnarray}
Adding these inequalities and dividing by three yields, for $N=4$,
\begin{equation}  A_Y \geq \frac{1}{N-1} \sum_{i<j}A_{ij}.  \end{equation}
The proof for general $N$ is similar.  In consequence, an intuitive
minimal-area argument selects the $\Delta$ law over the $Y$ law.

Now we consider some perturbative contributions.  Our results are a minor
sharpening of previous work\cite{c77,ff,wi} both for $N=3$ and for $N>3$,
showing how a large class of graphs with three-gluon (and sometimes four-
gluon) vertices vanishes identically.  In particular, those graphs formed
by attaching an $N$-leg connected gluon tree with $N-2$ three-gluon vertices
to $N$ quark lines in a baryon, such that only one gluon is attached to
each quark line, vanish; these are the ones with the topology of the
generic minimal Steiner surface, and we will call them Steiner graphs for
short.  Fig. 9, with the lines interpreted as gluon lines attached to the
numbered quarks, is an example for $N=4$.

It is well-known\cite{wi} that each individual Feynman graph in a baryon is
non-leading at large $N$, but the number of graphs is such that the sum
of all possible insertions of a given graph structure on the $N$ quark lines
is $O(N)$.  Moreover, the contribution of simple one-gluon exchange between two quarks is $1/(N-1)$ times one-gluon exchange in a $q\bar{q}$ loop. 
 But it was shown long ago\cite{c77,ff} that for $SU(3)$ the
graph of Fig. 10 vanishes identically, for symmetry reasons.  Let us
generalize this to any $N \geq 3$.  Consider the graph of Fig. 11, which
is characterized by having a gluon line joining quark lines 1 and 2 (for example), with
a single three-gluon vertex going somewhere else (neither to line 1 nor 2).
The group-theoretic factor of this graph is
\begin{equation}  \frac{1}{N!}\epsilon_{abc\cdots} \epsilon_{a'b'c' \cdots}
(\lambda^A)_{aa'}(\lambda^B)_{bb'} \cdots f_{ABC} \cdots  \end{equation}
where $\lambda^A,\lambda^B$ are group generators and $f_{ABC}$ the structure
constants.  It is clear that everything in this expression except the
structure constant is symmetric on the exchange of quark lines 1 and 2, or
equivalently on the exchange of $A$ and $B$; since $f_{ABC}$ is totally
antisymmetric, the whole expression vanishes.  

One can construct this way many graphs which vanish, in particular
Steiner graphs.  By collapsing a gluon line attached to two three-gluon
vertices, one can also find graphs with four-gluon vertices which also
vanish.  One can also show by the same symmetry argument that the
gluon-disconnected graphs containing a graph which would vanish by itself and
one which would not (see Fig. 12) also vanish.

A simple variant of the group structure in (53) allows one to calculate the
group-theoretic coefficient of a class of two-body graphs, especially
one-gluon exchange and radiative corrections thereto.  This is, as mentioned
in connection with equation (11), $1/(N-1)$ times the factors associated with
one-gluon exchange in the $q\bar{q}$ loop.  This same coefficient applies to
any two-body graph for which the matrix $M_{aa'}$ replacing $\lambda_{aa'}$ in (53) occurring on either quark line is traceless.

\section{Conclusions}

In this paper we have shown several things, based on the usual vortex
condensate and confinement by vortex-Wilson loop linking:
\begin{enumerate}
\item For all $SU(N),\;N>3$, the BWL area law is of the $\Delta$ form given 
in (45,46), and not of the $Y$ (or Steiner) form based on the generic 
minimum area.  
\item  The $\Delta$ area is weighted with $1/(N-1)$ in the BWL
area law relative to the unit weight of the Steiner area $A_Y$.  This is, as it must be, consistent with the argument that if $N-1$ quark lines are collapsed into an antiquark, the resultant area law is the $q\bar{q}$ law. 
\item  The weighted $\Delta$ area is less than the global minimum 
area $A_Y$ (and {\it a fortiori} any other area spanning the BWL), so even
an intuitive argument based only on minimization, that is, maximizing
$\langle BWL \rangle$, selects the $\Delta$ law.
\item Certain perturbative graphs with three-gluon vertices vanish by
a simple symmetry argument; these are graphs contributing to $3\cdots N$-body forces in a baryon.  Amopng them are the lowest-order graphs with the
topology of a Steiner surface cross-section.  
\item In the two-body sector, the sum of the (radiatively corrected) one-gluon potential and the
area law potential is $1/(N-1)$ times the corresponding $q\bar{q}$ potential.
At tree level, there is no perturbative contribution from graphs with the
topology of a cross-section of a Steiner surface.
\end{enumerate}

What can one make of these results?  Unfortunately, QCD theory (as opposed
to computer simulations) has not progressed to the point where it is possible
to distinguish the $\Delta$ law from the $Y$ law in a practical way, but of
course one optimistically hopes that this can be done some day.  There could
conceivably be some effect in using the $\Delta$ law in, e.g., fragmentation
models, but one is far from seeing such effects.  In the absence of meaningful
theoretical approaches to the baryon, it might still be useful to do more
precise computer simulations to verify the above claims.  The fact that
one-gluon perturbative exchange has a structure analogous to the $\Delta$ law
for areas, and no contribution to the structure analogous to the $Y$ (or generic minimal surface) law, may help to interpret these lattice calculations, where it
is of course impossible to separate perturbative and non-perturbative 
contributions to the BWL.

\newpage

\begin{center}{\bf ACKNOWLEDGEMENTS} \end{center}
\vspace{24pt}
$\; \; \;$ I thank Daniel Cangemi and John Garnett for useful conversations.
This work was supported in part by the National Science Foundation under
Grant PHY 9531023.

\newpage
\begin{center}{\bf Figure Captions}
\end{center}
\vspace{24pt}
Fig. 1.  The $SU(3)$ BWL.\\
Fig. 2.  Generic global minimal area of the BWL; the central line is the Steiner
line.\\
Fig. 3.  The $\Delta$ areas of the BWL; each pair of quark lines is spanned by
an area minimal for the loop formed by the two lines.\\
Fig. 4.  The three-bladed Stokes surface for the BWL non-Abelian Stokes'
theorem.  Also shown:  A vortex linked to line 1, passing through the
01 blade of the Stokes surface.\\
Fig. 5.  The same vortex passing through blades 02 and 03.\\
Fig. 6.  A Wilson loop (lines 1-8) and its decomposition into four
plaquettes (see Ref.\cite{fgk}).\\
Fig. 7.  A BWL composed of three square contours.\\
Fig. 8.  Dividing the BWL into plaquettes with a central line (the 0 line
of Fig. 4).\\
Fig. 9.  A cross-section of the $N=4$ BWL, with the quark-line cross-sections
shown as the numbered points, and the global minimum surface cross-section
as the lines; the points A, B, are cross-sections of Steiner lines.\\
Fig. 10.  Lowest-order three-gluon graph for an $SU(3)$ baryon.\\
Fig. 11.  A baryon graph for $SU(N)$ with a three-gluon vertex.\\
Fig. 12.  A gluon-disconnected baryon graph.  

\newpage
\epsfig{file=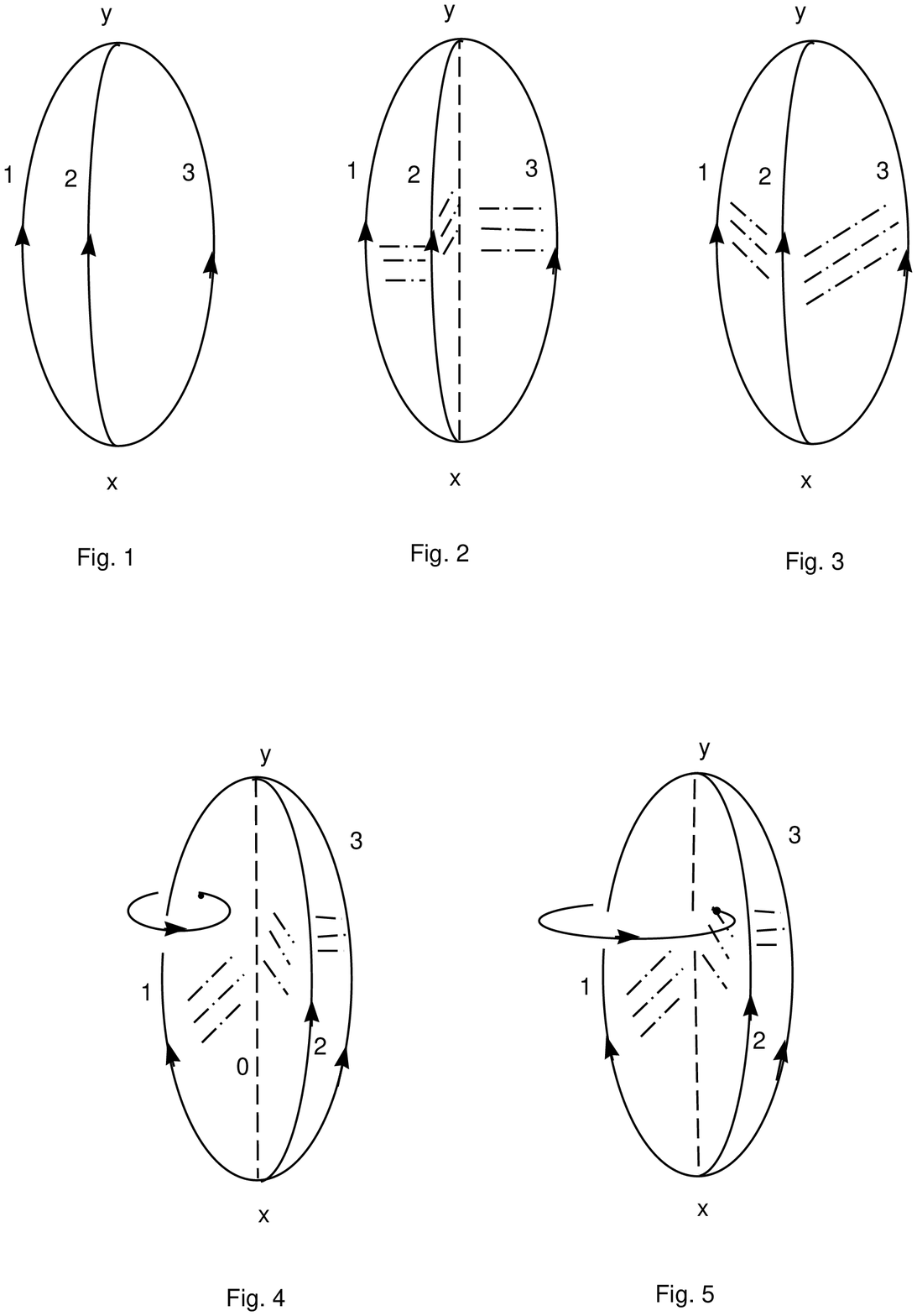,clip=,height=7in}
\newpage
\epsfig{file=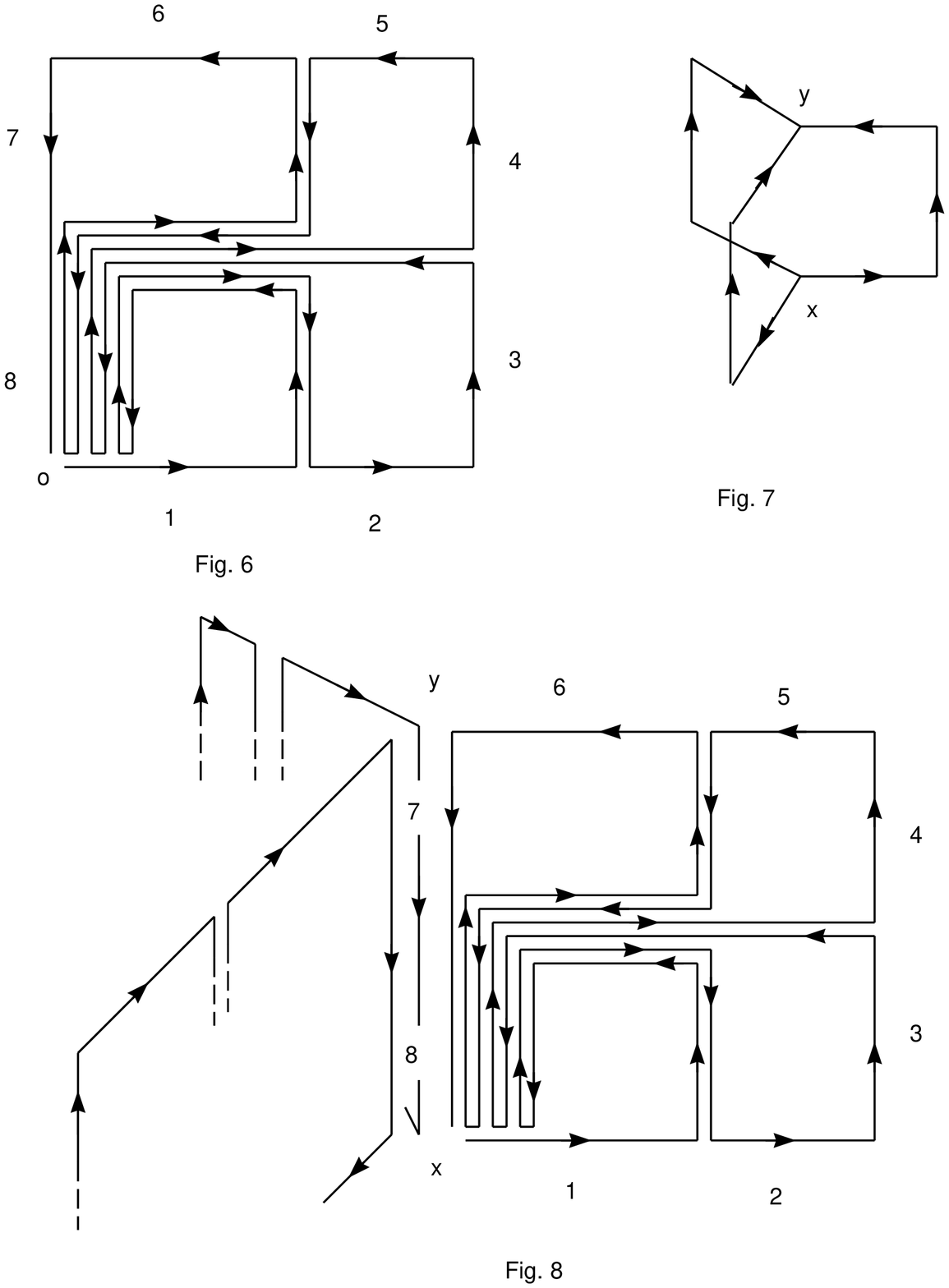,clip=,height=7in}
\newpage
\epsfig{file=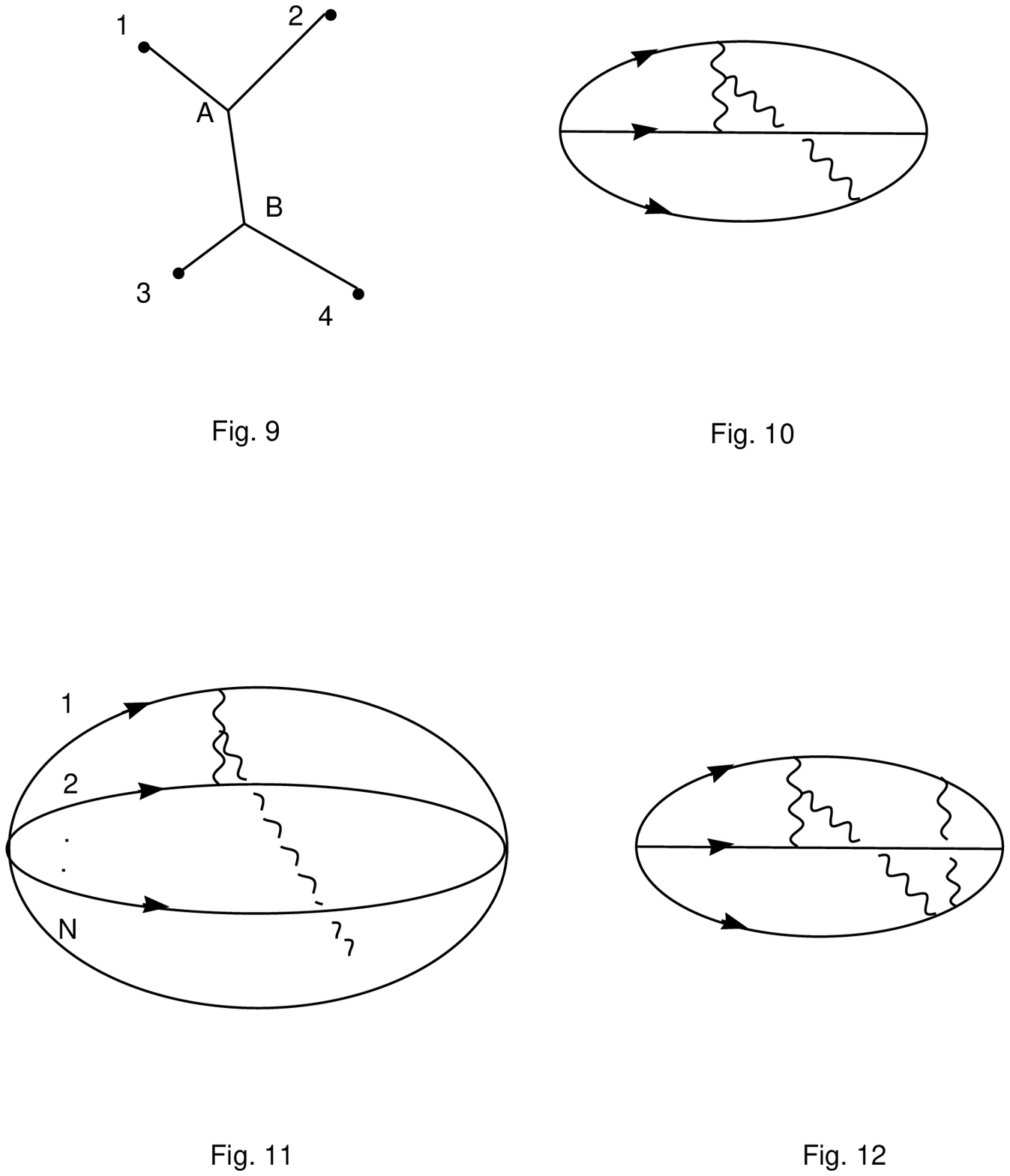,clip=,height=6in}

\end{document}